# JWST Reveals Bulge-dominated Star-forming Galaxies at Cosmic Noon

Chloë E. Benton[1] , Erica J. Nelson[1] , Tim B. Miller[2] , Rachel Bezanson[3] , Justus Gibson[1] , Abigail J Hartley[1] ,
Marco Martorano[4] , Sedona H. Price[3] , Katherine A. Suess[5,6] , Arjen van der Wel[4] , Pieter van Dokkum[7] ,
John R. Weaver[8] , and Katherine E. Whitaker[8,9]

[1] Department for Astrophysical and Planetary Science, University of Colorado, Boulder, CO 80309, USA
[2] Center for Interdisciplinary Exploration and Research in Astrophysics (CIERA), Northwestern University, 1800 Sherman Ave., Evanston, IL 60201, USA
[3] Department of Physics and Astronomy and PITT PACC, University of Pittsburgh, Pittsburgh, PA 15260, USA
[4] Sterrenkundig Observatorium, Universiteit Gent, Krijgslaan 281 S9, B-9000 Gent, Belgium
[5] Department of Astronomy and Astrophysics, University of California, Santa Cruz, 1156 High St., Santa Cruz, CA 95064, USA
[6] Kavli Institute for Particle Astrophysics and Cosmology and Department of Physics, Stanford University, Stanford, CA 94305, USA
[7] Department of Astronomy, Yale University, 52 Hillhouse Ave., New Haven, CT 06511, USA
[8] Department of Astronomy, University of Massachusetts, Amherst, MA 01003, USA
[9] Cosmic Dawn Center (DAWN), Niels Bohr Institute, University of Copenhagen, Jagtvej 128, København N, DK-2200, Denmark
Received 2024 August 13; revised 2024 September 18; accepted 2024 September 20; published 2024 October 14

## Abstract

Hubble Space Telescope imaging shows that most star-forming galaxies at cosmic noon—the peak of cosmic star formation history—appear disk-dominated, leaving the origin of the dense cores in their quiescent descendants unclear. With the James Webb Space Telescope's high-resolution imaging to 5 $\mu$m, we can now map the rest-frame near-infrared emission, a much closer proxy for stellar mass distribution, in these massive galaxies. We selected 70 star-forming galaxies with $10 < \log(M) < 12$ and $1.5 < z < 3$ in the CEERS survey and compare their morphologies in the rest-frame optical to those in the rest-frame near-IR. While the bulk of these galaxies are disk-dominated in 1.5 $\mu$m (rest-frame optical) imaging, they appear more bulge-dominated at 4.4 $\mu$m (rest-frame near-infrared). Our analysis reveals that in massive star-forming galaxies at z ∼ 2, the radial surface brightness profiles steepen significantly, from a slope of ∼0.3 dex$^{-1}$ at 1.5 $\mu$m to ∼1.4 dex$^{-1}$ at 4.4 $\mu$m within radii <1 kpc. Additionally, we find their total flux contained within the central 1 kpc is approximately 7 times higher in F444W than in F150W. In rest-optical emission, a galaxy's central surface density appears to be the strongest indicator of whether it is quenched or star-forming. Our most significant finding is that at redder wavelengths, the central surface density ratio between quiescent and star-forming galaxies dramatically decreases from ∼10 to ∼1. This suggests the high central densities associated with galaxy quenching are already in place during the star-forming phase, imposing new constraints on the transition from star formation to quiescence.

*Unified Astronomy Thesaurus concepts:* Galaxy evolution (594); Galaxy formation (595); Galaxy structure (622);
Galaxy bulges (578); Quenched galaxies (2016)

## 1. Introduction

Initial observations of galaxies at cosmic noon ($1 < z < 3$) show their morphologies are often irregular and clumpy in the rest-frame ultraviolet (e.g., S. P. Driver et al. 1995; K. Glazebrook et al. 1995; R. G. Abraham et al. 1996; C. J. Conselice et al. 2004; L. A. Moustakas et al. 2004; P. Cassata et al. 2005; J. M. Lotz et al. 2006). This evidence is thought to be an implication of galaxy growth driven by mergers during this epoch (C. J. Conselice et al. 2003; C. Papovich et al. 2005; J. M. Lotz et al. 2006; S. Ravindranath et al. 2006; G.-W. Fang et al. 2015; M. Park et al. 2022). This is due to the fact that regular morphology of galaxies can be considerably disrupted by mergers or ongoing interactions, leading to the formation of the distinct clumps and asymmetries seen in these galaxies. However, the installation of Wide Field Camera 3 (WFC3) on the Hubble Space Telescope (HST) and adaptive-optics-assisted near-infrared (NIR) integral field unit spectroscopy from ground-based telescopes enable sensitive, resolved rest-optical imaging and spectroscopy at these redshifts for the first time. NIR imaging reveals rest-optical light is smoother and more disk-dominated than the rest-UV light (L. E. Kuchinski et al. 2001; A. Rawat et al. 2009; J. S. Kartaltepe et al. 2012; S. Wuyts et al. 2012). NIR spectroscopy indicates rotation-dominated kinematics (K. L. Shapiro et al. 2008; N. M. Förster Schreiber et al. 2011, 2019; E. J. Nelson et al. 2013; E. Wisnioski et al. 2015) leading to a paradigm shift that most massive star-forming galaxies (SFGs) at early cosmic times are disk-dominated. The theory positing mergers as the primary catalyst for galaxy growth evolves into a new outlook, suggesting that gas accretion, rather than violent processes like mergers, is the prime suspect for gradual disk growth. In changing the observable rest-frame wavelength from optical to NIR, our understanding of galactic morphology evolves by revealing previously obscured features and dynamics within galaxies, which significantly alters our galaxy formation paradigm.

Previous work shows the stellar mass density within the central one kiloparsec ($\Sigma_1$) to be the best predictor of whether a galaxy is star-forming or quiescent (P. G. van Dokkum et al. 2014; G. Barro et al. 2017; K. E. Whitaker et al. 2017). However, it is not known how these dense cores (or bulges) form. Namely, we do not know how galaxies can transform their low central densities as SFGs to the high central densities of quiescent galaxies (QGs). Many studies demonstrate that SFGs are disk-dominated while QGs are bulge-dominated (e.g., S. Wuyts et al. 2011; E. Cheung et al. 2012; A. F. L. Bluck et al. 2014; N. Mandelker et al. 2014; P. Lang et al. 2014; P. G. van Dokkum et al. 2014; M. Mosleh et al. 2017; K. E. Whitaker et al. 2017; Y. Luo et al. 2020; J. Fensch & F. Bournaud 2021; H. Gao et al. 2022; M. Park et al. 2022; J. Puschnig et al. 2023). Yet, it remains inconclusive how and when this transition occurs. The prominent explanation is that there has to be a radical physical process associated with simultaneous quenching for this structure change (e.g., A. Toomre 1977; J. Barnes et al. 1988; L. Hernquist 1992, 1993; M. Noguchi 1999; V. Springel et al. 2005; A. Dekel et al. 2009; P. F. Hopkins et al. 2010; J. Weaver et al. 2018). James Webb Space Telescope's (JWST's) unparalleled high spatial resolution in the infrared allows us, for the first time, to probe the distribution of older and dustier stellar populations in the rest-frame NIR. This has the potential to revolutionize our understanding of galaxy structure and evolution, much like the shift from rest-UV to rest-optical did.

The rest-frame NIR emission is more reliable for determining the underlying stellar mass distribution compared to the rest-optical, as it exhibits smaller mass-to-light (M/L) gradients influenced by dust and age (E. F. Bell & R. S. de Jong 2001; G. Kauffmann et al. 2003a; E. F. Bell et al. 2003). While rest-frame optical emission better traces the bulk of a galaxy's stellar mass than rest-frame UV emission, multiple studies using HST imaging find that the M/L ratio varies throughout the galaxy, particularly as a function of radius (Y. Guo et al. 2011; D. Szomoru et al. 2013; S. Wuyts et al. 2013; F. S. Liu et al. 2016; W. Wang et al. 2017; K. A. Suess et al. 2019a, 2019b; T. B. Miller et al. 2023). Previously, no facility could measure the detailed structure of galaxies at $z \sim 2$ in the rest-NIR. Now, with JWST's incredible sensitivity and resolution in the infrared, we can, for the first time, accurately observe the rest-frame NIR morphology of galaxies at cosmic noon.

Our study compares the morphology of galaxies at 1.50 and 4.44 $\mu$m using JWST; 1.50 $\mu$m (F150W) is comparable to the reddest HST filter (F160W), corresponding to rest-optical for galaxies at cosmic noon. We expect the 4.44 $\mu$m (F444W) emission, corresponding to rest-frame NIR, to be a better tracer of stellar mass because it is less affected by dust extinction and more sensitive to older, cooler stars, providing a more accurate representation of the overall stellar mass distribution. So far, JWST observations show that galaxies with higher stellar masses ($\sim 10^{11} M_\odot$) observed at 4.44 $\mu$m have up to 30% smaller sizes than in 1.5 $\mu$m emission (K. A. Suess et al. 2022; A. van der Wel et al. 2024). Further investigation shows that this may be caused by galaxies having color gradients (i.e., negative color gradients) that are driven primarily by dust or age (T. B. Miller et al. 2022; S. H. Price et al. 2023). On the other hand, M. Martorano et al. (2023) show that while light profiles in the UV/optical can be influenced by young stars and dust absorption, and the NIR more accurately represents the stellar mass distribution, there is a significant absence of strong wavelength dependence of the Sérsic index for both SFGs and QGs over redshifts $0 < z < 3$.

In this Letter, we investigate the radial surface brightness profiles and central surface flux densities of massive galaxies at $1.5 < z < 3$ as inferred from imaging in rest-optical compared to rest-NIR. In Sections 2 and 3, we provide a brief analysis of imcascade and the surface brightness profiles for each filter. In Section 4, we discuss the implications of these radial profiles with respect to their central surface densities ($R < 1$ kpc).

Lastly, in Section 5, we present our conclusion and ultimately, the ramifications these findings have on morphological evolution. We assume a flat $\Lambda$CDM cosmology with $H_0 = 70$ km s$^{-1}$ Mpc$^{-1}$ and $\Omega_M = 0.3$. All radii are referred to along the semimajor axis and all magnitudes are reported on the AB system (J. B. Oke 1974).

## 2. Data

We use the Cosmic Evolution Early Release Science Survey (CEERS) program, ERS Program 1345 (S. L. Finkelstein et al. 2017; S. Finkelstein et al. 2023). CEERS covers $\sim 100$ arcmin$^2$ of the Extended Groth Strip HST legacy field using NIRCam, MIRI, and NIRSpec (G. Yang et al. 2021; M. B. Bagley et al. 2023). For this project, we use NIRCam imaging in the F150W and F444W filters. G. Brammer (2023) conducts the data reduction, coadding, and aligning of the exposures using the public software package grizli (see also G. Brammer & J. Matharu 2021; I. Labbe et al. 2023). Briefly, grizli masks imaging artifacts, subtracts the sky background, and aligns the images to stars from Gaia-DR3.

All fits are performed on the $\sim 0''03$ drizzled images, which use empirical point-spread functions (ePSFs) from Z. Ji et al. (2023), see also K. A. Suess et al. (2023). The ePSFs were created using the EPSFBuilder in PHOTUTILS, which adopts the method from J. Anderson & I. R. King (2000). This empirical method iteratively solves the centroids and fluxes of a list of input point sources, and then stacks them together.

We select galaxies from the public redshift and stellar population catalogs provided by the 3D-HST survey (G. B. Brammer et al. 2012; R. E. Skelton et al. 2014, v4.1.5). The 3D-HST is a 248-orbit extragalactic treasury program with HST furnishing NIR spectroscopy over the five CANDELS fields. The redshifts are derived by fitting the photometry and the 2D G141 spectrum with a modified version of the EAzY code (G. B. Brammer et al. 2008). The galaxy stellar masses and rest-frame colors are fit by utilizing the stellar population synthesis modeling of the photometry using the FAST code (M. Kriek et al. 2009).

## 3. Analysis

We select galaxies with coverage in both F150W and F444W in CEERS (S. L. Finkelstein et al. 2017), redshifts $1.5 < z < 3$, $\log(M_*/M_\odot) > 10.5$, and characterize them as star-forming according to their rest-UVJ colors in the 3D-HST catalog (e.g., see K. E. Whitaker et al. 2012). The four criteria result in an initial sample of 79 SFGs, as shown in Figure 1. We exclude nine galaxies that are flagged by imcascade as ill-converged fits. This leaves a sample of 70 galaxies to analyze. We confirm that the exclusions do not bias our results. Both the T-Test and Kolmogorov–Smirnov test indicate that the mass distribution of our sample galaxies is similar to that of the excluded ones. This supports the conclusion that the exclusion did not introduce a significant bias. Additionally, the mean mass of the excluded galaxies is comparable to that of our sample, further suggesting that the exclusion likely did not introduce bias with respect to average mass.

We provide images of four example galaxies in Figure 2. It is clear that galaxies are more bulge-dominated in F444W light than in F150W light. We quantify these trends by constructing PSF-corrected radial surface brightness profiles. We use the open-source software package imcascade to model the light distribution of galaxies, following T. B. Miller & P. van Dokkum (2021). `imcascade` uses a Bayesian implementation of the Multi-Gaussian Expansion method to model galaxies as a mixture of Gaussians, with a Gaussian decomposition of the PSF to enable analytic convolution (T. B. Miller et al. 2023). The fitting process employs a fixed set of Gaussian widths, and `imcascade` then determines the weights for each Gaussian component.

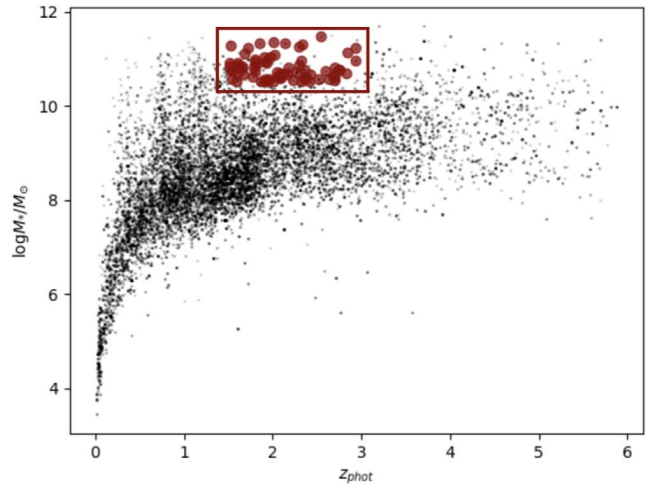

**Figure 1.** Redshifts and stellar masses for the galaxies in our sample in red. Our galaxies have $1.5 < z < 3$ and $\log(M_*/M_\odot) \sim 10^{10}$–$10^{12}$. The black dots show the CEERS AEGIS field galaxies (G. B. Brammer et al. 2012; R. E. Skelton et al. 2014).

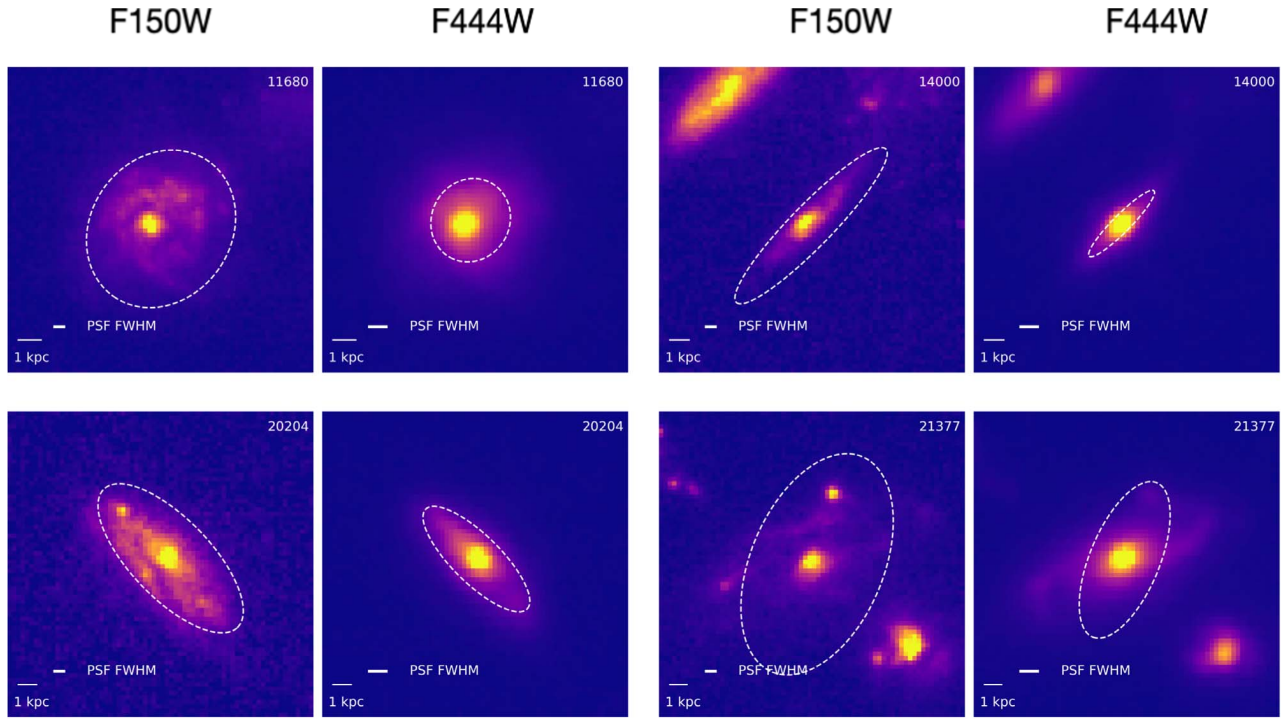

**Figure 2.** Comparison of galaxy images in the F150W and F444W filters produced by JWST. The white ellipse shows the effective radius of each galaxy's fits. Each galaxy's PSF FWHM is also shown. While the bulges of these $1.5 < z < 3$ galaxies appear small in the F150W, we see their dominance when looking at the F444W image.

The results from the `imcascade` optimization and Bayesian fitting method produces posterior distributions of the structural parameters (e.g., center position, position angle, axis ratio, etc.), which we use to construct PSF-corrected radial profiles. We generate a PSF-corrected galaxy model by summing the constituent Gaussians without convolving them with the PSF and then adding back the fit residuals. By incorporating these residuals, we account for discrepancies between the model and observed data. We note that while the residuals are not PSF-corrected, incorporating them is crucial because a galaxy's structure cannot be fully captured by a sum of Gaussians alone. Including the residuals helps account for additional features of the galaxy, leading to a more realistic model and mitigating potential biases in our analysis, even if some PSF smearing effects persist. A similar method has often been applied using `GALFIT` (e.g., C. Y. Peng et al. 2002, 2010, 2011; E. J. Nelson et al. 2013; K. A. Suess et al. 2019b; B. A. Pastrav 2020; C. Dewsnap et al. 2023;

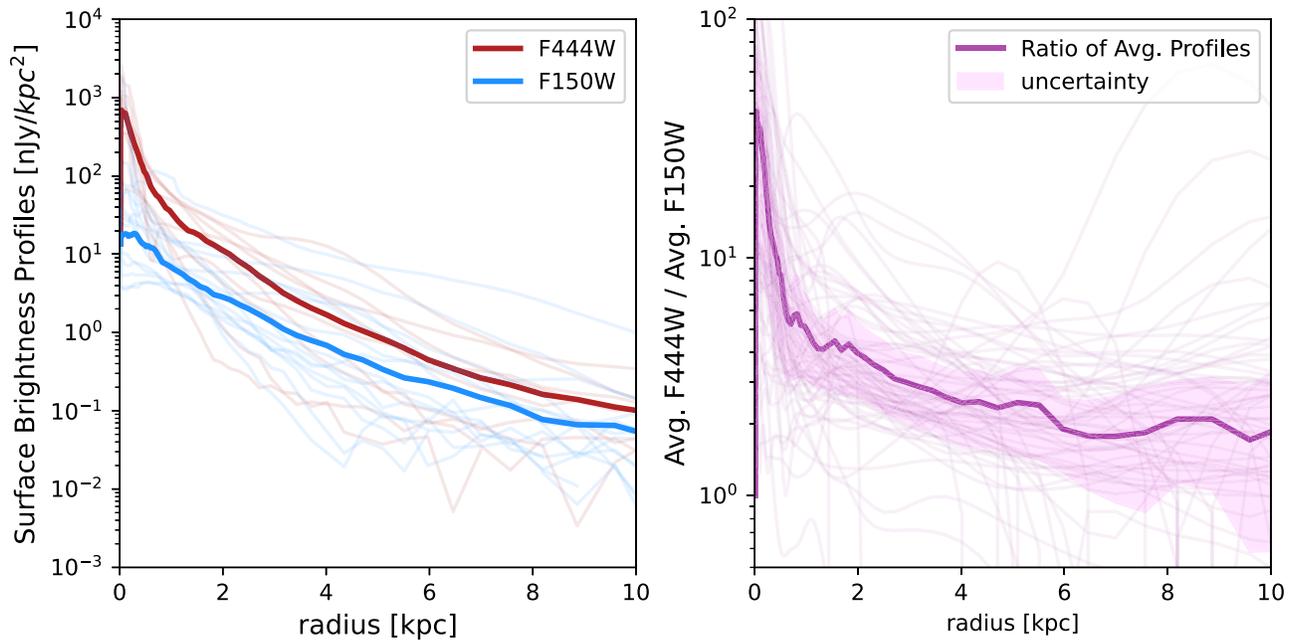

**Figure 3.** The left panel shows the radial surface brightness profiles of our SFGs given in F150W and F444W. The light colored lines show the surface brightness profiles of individual galaxies of each filter, the dark their median. Overall, F444W has a much steeper surface brightness profile than the F150W. For the right panel, the shaded region represents the uncertainty in the ratio of the averaged profiles. The light colored lines represent the ratio of the filters (F444W/F150W) for each individual profile. As we expect, there is more variation at the outskirts of these galaxies.

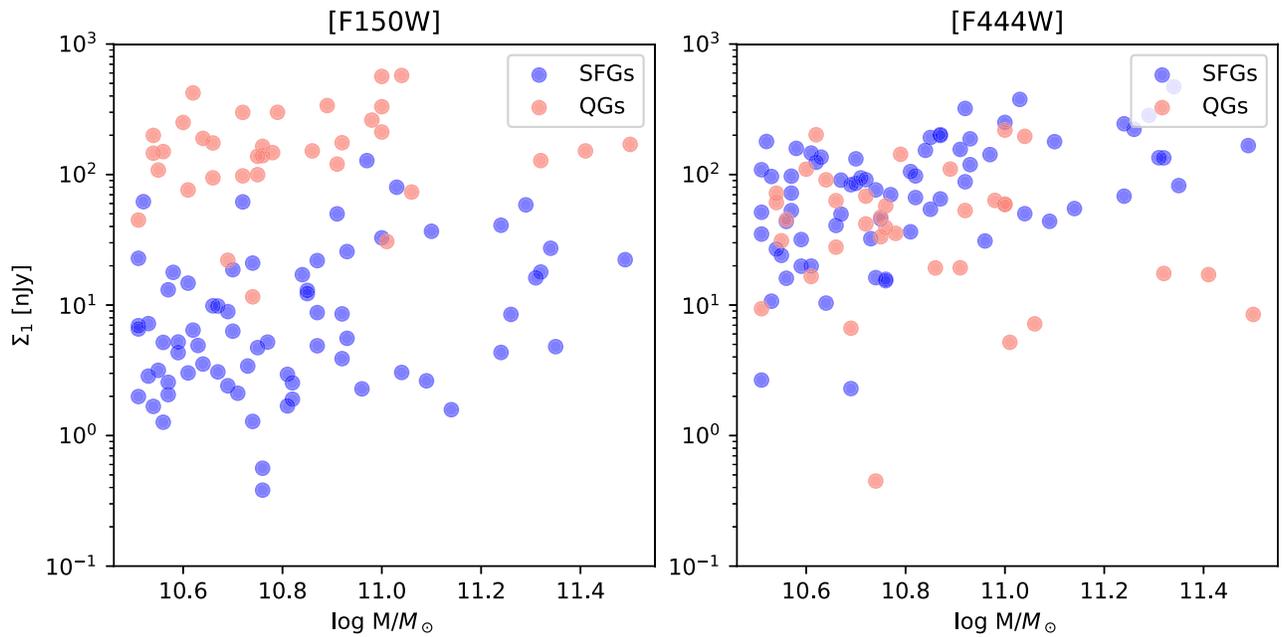

**Figure 4.** A comparison of the central flux of galaxies in F150W (rest-frame $V$ band; left) vs. F444W (rest-frame $J$ band; right). $\Sigma_1$ is the flux contained within the central kiloparsec. Here, $\Sigma_1$ represents central flux surface density instead of stellar mass density. We find that galaxies measured in F444W overall have higher central flux than in F150W.

L. Tortorelli & A. Mercurio 2023). We use `imcascade` instead of `GALFIT` because it is nonparametric such that the results are not biased by parametric fits (T. B. Miller & P. van Dokkum 2021).

We derive the radial surface brightness profiles using elliptical apertures through `photutils` on the PSF-corrected models. We use the best-fit axis ratio, center, and position angle produced by `imcascade` for the elliptical apertures. This returns the summed flux within the given apertures. We get the radial surface brightness profiles by dividing the sum of the values within the aperture by the exact analytical area of the aperture shape, then normalizing the aperture profiles to the innermost aperture. This ensures that the profile values are relative to the first aperture's profile, and it scales all the fluxes by the flux in the innermost aperture.

## 4. Results

Figure 3 shows the PSF-corrected radial surface brightness profiles for our sample of 70 massive SFGs at $1.5 < z < 3$. We show the individual profiles (light colored) and median profiles (dark colored) in both F150W and F444W—first panel. While there is variety among the individual profiles, the median slope at 4.44 $\mu$m is 5 times steeper within $<$1 kpc compared to 1.50 $\mu$m, indicating much more centralized flux in F444W than in F150W. Explicitly, the median profile at 1.50 $\mu$m within radii $<$ 1 kpc has a slope of $\sim$0.3 dex$^{-1}$, while the median profile at 4.44 $\mu$m has a slope of $\sim$1.4 dex$^{-1}$. If we interpret a steeper central slope, i.e., concentrated flux, as reflecting a stronger bulge, then the galaxies observed at 4.44 $\mu$m have a significantly more prominent bulge. Here, we have a closer representation of stellar mass through the F444W filter.

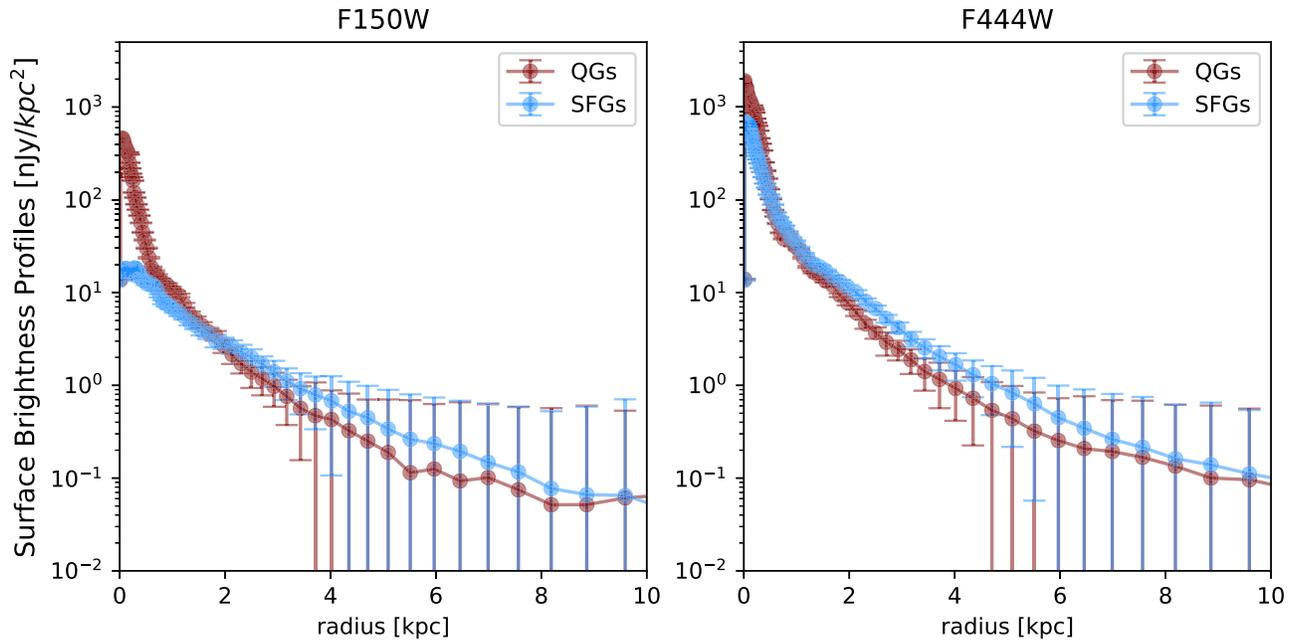

**Figure 5.** The average radial brightness profile for star-forming galaxies and quiescent galaxies at filters F150W and F444W. Right: we note that both galaxies have dense cores, with the peaks being at relative magnitudes. Left: at shorter wavelengths, the peak of the star-forming galaxies is a magnitude lower than the quiescent galaxies. This offset is to be expected since we assume the F150W filter to be an inconsistent tracer of stellar mass.

To quantify the trend of more centrally concentrated light in F444W than F150W, we compute the total flux contained within the central kiloparsec ($\Sigma_1$) by integrating the profiles to 1 kpc. Figure 4 shows the $\Sigma_1$ values relative to stellar mass for both wavelengths, revealing that most of these massive SFGs have significantly higher central flux fractions in F444W than in F150W. For this study, we use $\Sigma_1$ as an indicator of central flux surface density rather than stellar mass density. Our analysis indicates that SFGs have, on average, about 7 times more central flux in the rest-frame NIR than in the rest-frame optical. This suggests that SFGs have much denser cores than previously thought based on their rest-frame optical emission. This has significant impacts for our understanding of galaxy structural transformation, i.e., the dense cores of galaxies are built up while they are star-forming. These dense cores are significant as they are closely linked to the quenching processes in galaxies.

## 5. Discussion

Previous work finds that dense cores are intimately intertwined with galaxy quenching: the presence of a dense core is one of the best predictors of quiescence in galaxies (G. Kauffmann et al. 2003b, 2006; D. Schiminovich et al.

2007; E. F. Bell 2008; E. Cheung et al. 2012; J. J. Fang et al. 2013; P. Lang et al. 2014; P. G. van Dokkum et al. 2014; L. E. Abramson et al. 2016; G. Barro et al. 2017; K. E. Whitaker et al. 2017). When a dense core is present, a galaxy is more likely to be quenched. However, it remains ambiguous how this structural transformation from disk-dominated to bulge-dominated occurs during the transition from star-forming to QGs. Fascinatingly, our findings show that SFGs already have much more prominent dense cores than previously thought based on optical imaging. The immediate question now is how the central densities compare, given that we can observe both populations in the rest-NIR light.

To investigate this, we select QGs based on their rest-UVJ colors in the same mass and redshift range as the SFGs in our sample. We measure the PSF-corrected surface brightness profiles and central flux surface densities for the QGs using the method discussed in Section 3 for our SFGs. Figures 4 and 5 show our results. Figure 5 shows $1.5 < z < 3$ QGs have steeper slopes, on average, within the central 1 kpc compared to their star-forming counterparts in F150W (left panel). Our results are consistent with a key result in M. Martorano et al. (2023), which presents, through Sérsic measurements, that the radial profiles of SFGs and QGs are indeed different at shorter NIR wavelengths ($\sim$1.1 $\mu$m). Our key result, however, is that in F444W, the central surface density within 1 kpc shows a dramatic decrease in difference, narrowing to approximately a factor of $\sim$1.4 (right panel). This means the QGs and SFGs measured at 4.44 $\mu$m have much more similar central surface densities and profiles. Figure 4 visually quantifies this by plotting the total flux contained within the central kiloparsec ($\Sigma_1$) relative to their stellar mass for QGs and SFGs for each filter. $\Sigma_1$ values are reflective of central flux measurements and may not accurately represent the underlying mass profiles, as radial variations in the M/L ratio may still need to be accounted for. Where there is an offset between the central flux of QGs and SFGs in F150W by a factor of $\sim$10, it dramatically reduces in F444W, meaning both QGs and massive SFGs have dense cores in F444W.

Our observations indicate that SFGs at $1.5 < z < 3$ possess dense cores, as evidenced by high central surface densities measured in the rest-frame NIR. This suggests that the transition from disk-dominated to bulge-dominated structures does not necessarily require a radical structural transformation at fixed mass. In other words, the development of dense central regions, typically associated with bulges, occurs prior to the quenching of star formation. Furthermore, our findings suggest a potentially missing phase in galaxy evolution. While high central densities have been strongly associated with quiescence, these dense cores must first form. We observe massive SFGs with prominent dense cores, indicating that a population of bulges-in-formation exists. This implies that high central surface density does not immediately lead to the cessation of star formation. This discrepancy could be due, in part, to the lack of rest-frame NIR photometry in previous high-$z$ $\Sigma_1$ studies, which may have overlooked dense core formation in actively SFGs. Thus, it is possible for galaxies to quench without undergoing a rapid and dramatic change in their overall morphology. The increase in central flux density in SFGs observed in the rest-frame NIR compared to the rest-frame optical could indicate the presence of either old, red stars or dust-obscured young stars (e.g., T. B. Miller et al. 2022). In the former case, the dense centers may have formed at early cosmic times as relics of galaxy formation in a dense universe, remaining hidden until now. In the latter case, these dense centers may be actively building through ongoing star formation (E. J. Nelson et al. 2016, 2019; T. B. Miller et al. 2022).

To summarize, our findings show that while SFGs at $z \sim 2$ are generally disk-dominated in the rest-frame optical, they appear much more bulge-dominated in the rest-frame NIR, with dense cores typically associated with QGs already in place. However, there remain open questions regarding the exact processes responsible for the formation of these dense cores and the timeline of their development. Specifically, it is unclear whether the high central surface densities are primarily due to dust-obscured star formation or represent an established bulge structure formed through past events. Addressing these questions will require constraining the spatially resolved stellar populations within galaxies, particularly through direct measurements of their star formation histories at cosmic noon. These methods will provide a clearer understanding of the evolutionary pathways of massive galaxies and the underlying mechanisms that drive their structural transformation and quenching.

## Acknowledgments

Support for this work was provided by NASA through grant HST-AR-16146 and JWST-GO-01810 awarded by the Space Telescope Science Institute, which is operated by the Association of Universities for Research in Astronomy, Inc., under NASA contract NAS 5-26555. These observations are associated with program JWST-ERS-1345. The data products presented herein were retrieved from the Dawn JWST Archive (DJA).

*Facility:* JWST (NIRCam).

*Software*: astropy (Astropy Collaboration et al. 2013, 2018, 2022), imcascade (T. B. Miller & P. van Dokkum 2021).

## ORCID iDs

Chloë E. Benton https://orcid.org/0000-0001-5378-9998
Erica J. Nelson https://orcid.org/0000-0002-7524-374X
Tim B. Miller https://orcid.org/0000-0001-8367-6265
Rachel Bezanson https://orcid.org/0000-0001-5063-8254
Justus Gibson https://orcid.org/0000-0003-1903-9813
Abigail I Hartley https://orcid.org/0000-0002-5891-1603
Marco Martorano https://orcid.org/0000-0003-2373-0404
Sedona H. Price https://orcid.org/0000-0002-0108-4176
Katherine A. Suess https://orcid.org/0000-0002-1714-1905
Arjen van der Wel https://orcid.org/0000-0002-5027-0135
Pieter van Dokkum https://orcid.org/0000-0002-8282-9888
John R. Weaver https://orcid.org/0000-0003-1614-196X
Katherine E. Whitaker https://orcid.org/0000-0001-7160-3632